\input mtexsis
\paper
\singlespaced
\widenspacing
\twelvepoint
\Eurostyletrue
\sectionminspace=0.1\vsize
%definitions
\def\discrete{{\bf Z }_2} 
\def\discr{{\bf S }_2}

\def\yo1{{{f_\pi}^2}}

\def\oneht{\textstyle{1\over 2} }
\def\onehtsq{\textstyle{1\over{\sqrt{2}}} }

\def\sss{\scriptscriptstyle}

\def\ql{{ q_{\sss L} }}
\def\qr{{ q_{\sss R} }}
\def\pl{{ p_{\sss L} }}
\def\pr{{ p_{\sss R} }}
\def\bql{{ {\bar q}_{\sss L} }}
\def\bqr{{ {\bar q}_{\sss R} }}
\def\bpl{{ {\bar p}_{\sss L} }}
\def\bpr{{ {\bar p}_{\sss R} }}
%references
\referencelist
\reference{beane} S.R.~Beane, Maryland PP\#97-075, {{\tt hep-ph/9706246}}
\endreference
\reference{pdg} Particle Data Group, \journal Phys. Rev.D;54, (1996)
\endreference
\reference{thooft}  G.~'t Hooft,
                    in {\it Recent Developments in Gauge Theories},
                    Proc. of the NATO ASI, Carg{\`e}se 1979,
                    ed. G.'t Hooft {\it et al} (Plenum, 1980)
\endreference
\reference{mended}  S.~Weinberg, \journal Phys. Rev. Lett.;65,1177 (1990);
           {\it ibid}, 1181
\endreference
\reference{alg}  S.~Weinberg, \journal Phys. Rev.;177,2604 (1969)
\endreference
\reference{*alga} See also, F.J.~Gilman and H.~Harari,
                  \journal Phys. Rev.;165,1803 (1968)
\endreference
\reference{*algb} B.~Zumino, in {\it Theory and Phenomenology in Particle
Physics}, 1968 International School of Physics `Ettore Majorana'
(Academic Press, New York, 1969) p.42
\endreference
\reference{*algc} S.~Weinberg, in {\it Lectures on Elementary Particles and
Quantum Field Theory}, edited by Stanley Deser {\it et al}, 
(MIT Press, Cambridge, MA, 1970) p.285
\endreference 
\reference{*algd}  R.~de Alfaro, S.~Fubini, G.~Furlan, and G.~Rossetti, 
            {\it Currents in Hadron Physics}, (North-Holland, Amsterdam, 1973)
\endreference
\reference{wilson}  K.G.~Wilson and D.G.~Robertson,  OSU-NT-94-08, {{\tt hep-th/9411007}}
\endreference
\reference{*wilsona}  D.~Mustaki, {{\tt hep-ph/9410313}}
\endreference
\reference{sfsr}  S.~Weinberg, \journal Phys. Rev. Lett.;18,507 (1967)
\endreference
\reference{ksrf} K.~Kawarabayashi and M.~Suzuki,
                    \journal Phys. Rev. Lett.;38,883 (1966)
\endreference
\reference{*ksrfa} Riazuddin and Fayyazuddin,
                    \journal Phys. Rev.;147,1071 (1966)
\endreference
\reference{krauss} See, for instance, L.M.~Krauss and F.~Wilczek, 
      \journal Phys. Rev. Lett.;62,1221 (1989)
\endreference
\reference{bala}  A.P.~Balachandran and S.~Vaidya, \journal 
Phys. Rev. Lett.;78,13 (1997);\hfill\break SU-4240-653, {{\tt hep-ph/9612053}}
\endreference
\endreferencelist
% title page
\titlepage
\obeylines
\hskip4.8in{DOE/ER/40762-120}
\hskip4.8in{U.ofMd.PP\#97-113}
\hskip4.8in{hep-ph/9706250}\unobeylines
%\line{ \hfill \TeXsis\ \fmtversion}     % a banner of sorts
\title
Quantum Discrete Symmetry and the Strong CP Problem
\endtitle
\author
Silas R.~Beane

Department of Physics, Duke University
Durham, NC 27708-0305

{\it and}

Department of Physics, University of Maryland\Footnote\dag{Present Address.\hfill}
College Park, MD 20742-4111
\vskip0.1in
\center{{\it sbeane@pion.umd.edu}}\endcenter
\endauthor

\abstract
\singlespaced
\widenspacing

We study a $2N$-flavor effective theory of $N$-flavor QCD.  With the
axial anomaly accounted for in the effective theory by a 't Hooft
interaction, only QCD conserved currents survive. However, there is a
residual discrete symmetry with interesting properties. With
non-vanishing quark masses, this $\discr$ symmetry is broken unless
${\bar\theta}$=$0$ or $\pi$. We further show that there is a sense in
which hadrons in the effective theory fall into $\discr$ multiplets.
Surprisingly, predictions of this multiplet structure in the
four-flavor effective theory are in good agreement with experiment and
have been found previously by imposing Regge asymptotic constraints on
pion-hadron scattering amplitudes.

\endabstract 
\endtitlepage
\vfill\eject                                     % new page
% introduction
\superrefsfalse
\singlespaced
\widenspacing

%%%%%%%%%%%%%%%%%%%%%%%%%%%%%%%%%%%%%%%%%%%%%%%%%%%%%%%%%%%%%%%%%%%
\vskip0.1in
\noindent {\twelvepoint{\bf 1.\quad Introduction}}
\vskip0.1in

In this letter we consider the possibility that there exists an
effective theory with $2N$ flavors that describes the same low-energy
physics as QCD with $N$ flavors\ref{beane}.  The $2N$-flavor effective
theory can be sensible only if it has the same continuous global
symmetries as QCD since more or fewer conserved currents would almost
certainly do violence to the particle data group. We show that when
the axial anomaly is accounted for in the effective theory by a 't
Hooft interaction, only QCD conserved currents survive. However, there
remains a discrete symmetry which interchanges quark multiplets of
opposite chirality. This $\discr$ symmetry is the focus of this
letter. We demonstrate that if $\discr$ is unbroken, P and CP are also
unbroken. Therefore, the $2N$-flavor effective theory does not have a
strong CP problem in the usual sense. This immediately raises the
question of whether there are other testable consequences of
$\discr$. We show that $\discr$ makes several familiar and successful
predictions which were obtained long ago by assuming soft asymptotic
behavior in forward pion-hadron scattering.

Our letter is organized as follows.  In section 2, as a warm up, we
review the strong CP problem in $N$-flavor QCD with a 't Hooft
interaction.  In section 3 we construct the $2N$-flavor effective
theory and show that the effective theta angle, $\bar\theta$, is a
natural parameter in the sense that when it vanishes there is an
enhanced $\discr$ symmetry. In section 4 we show that $\discr$ has
algebraic consequences that have been found previously by imposing
Regge asymptotic constraints on pion-hadron scattering amplitudes.
The predictions of $\discr$ are in good agreement with
experiment. Finally, in section 5 we summarize and comment on the
significance of our results.

%%%%%%%%%%%%%%%%%%%%%%%%%%%%%%%%%%%%%%%%%%%%%%%%%%%%%%%%%%%%%%%%%%%
\vskip0.1in
\noindent {\twelvepoint{\bf 2.\quad The Strong CP Problem in $N$-flavor QCD}}
\vskip0.1in

In order to provide contrast for our results, we first give a simple
review of the strong CP problem in $N$-flavor QCD. We do not include
explicit gauge fields and color indices are suppressed. The matter
content of QCD consists of $N$ Dirac fermions assembled into the
vector, $q$, in the fundamental representation of $SU({{{N}}})$, which
transforms with respect to $SU({{{N}}})_L\times SU({{{N}}})_R$ as

\offparens
$$\eqalign{
&({{{N}}},1):\qquad\ql \rightarrow L\ql  \cr
&(1,{{{N}}}):\qquad\qr \rightarrow R\qr .\cr}
\EQN qpal2qcd$$
\autoparens
The most general $SU({{{N}}})_L\times SU({{{N}}})_R$
invariant free lagrangian one can build with $q$ is

\offparens
$$
{\cal L}_0=
{\bql} i\slashchar{\partial} {\ql} +
{\bqr} i\slashchar{\partial} {\qr}= 
{\bar q}i\slashchar{\partial} q .
\EQN qpal3qcd$$\autoparens
This free lagrangian also admits $U(1)_B$ and $U(1)_A$
transformations. We therefore add a $U(1)_A$ violating quark
interaction to take into account the effect of the axial anomaly.
Consider the $U(1)_A$ violating, $SU({{{N}}})_L\times
SU({{{N}}})_R\times U(1)_B$ preserving 't Hooft interaction

\offparens
$$
{\cal L}''({\bar\theta} )=  
-{\bar\kappa}\{ e^{i{\bar\theta}}\;det\; 
{\bar q}(1-{\gamma_5}){q} +
e^{-i{\bar\theta}}\;det\; 
{\bar q}(1+{\gamma_5}){q} \}, 
\EQN u1expbrqcd$$\autoparens
where the determinant acts on $SU({{{N}}})$ matrices and $\bar\kappa$
is a parameter of mass dimension $4-3{{{N}}}$. We have included a P
and CP violating phase, ${\bar\theta}$, which includes pure QCD
effects as well as quark mass matrix effects; i.e.
${\bar\theta}$=${\theta_{\sss QCD}}$+${\theta_{\sss EW}}$.

In the absence of explicit chiral symmetry breaking effects the field
redefinition, $q\rightarrow e^{i{\bar\theta}{\gamma_5}/2N}q$, removes
${\bar\theta}$ from the lagrangian, and then P and CP are manifest
discrete symmetries of the theory.  

Assuming $N$ degenerate flavors, we can include an explicit chiral
symmetry breaking mass term, $-{m_q}{\bar q}{q}$, and again perform a
field redefinition which removes ${\bar\theta}$ from the 't Hooft
interaction.  However, now ${\bar\theta}$ cannot be removed from the
lagrangian.  Generally, redefining a single quark flavor is sufficient
to remove ${\bar\theta}$ from the 't Hooft interaction. It is
therefore sufficient that one quark flavor be massless in order to
render ${\bar\theta}$ unphysical.  For small ${\bar\theta}$, ${\bar
q}{q}\rightarrow {\bar q}{q} +i({\bar\theta}/N){\bar q}{\gamma_5}{q}$,
which induces the P and CP violating operator,

$$
-i{{m_q}\over N}\,{\bar\theta}\,{\bar q}{\gamma_5}{q}.
\EQN ebs2qcd$$
If nonvanishing, this operator will contribute to low-energy physics,
e.g.  the neutron electric dipole moment. Experimentally one finds
$|{\bar\theta}|\leq 10^{-9}$\ref{pdg}.  Small parameters are considered
natural if their vanishing implies enhanced symmetry\ref{thooft}.  The
fact that ${\bar\theta}$ is so small and yet the symmetry of the
standard model is not increased by taking ${\bar\theta}$=$0$ is known
as the strong CP problem. It is true that CP becomes a good symmetry
of the strong interactions when ${\bar\theta}$=$0$. However, CP is
violated by the weak interactions, as observed in kaon decays, and is
therefore not a symmetry of the standard model.

%%%%%%%%%%%%%%%%%%%%%%%%%%%%%%%%%%%%%%%%%%%%%%%%%%%%%%%%%%%%%%%%%%%
\vskip0.1in
\noindent {\twelvepoint{\bf 3.\quad Absence of a Strong CP Problem in a $2N$-flavor Effective Theory of QCD}}
\vskip0.1in

We now repeat the analysis in a $2N$-flavor low-energy effective
theory of $N$-flavor QCD.  The matter content of the effective theory
consists of Dirac fermions, $q$ and $p$, each in the fundamental
representation of $SU({{{N}}})$, which transform with respect to
$SU({{{N}}})_L\times SU({{{N}}})_R$ as

\offparens
$$\eqalign{
&({{{N}}},1):\qquad\ql \rightarrow L\ql \qquad\pr \rightarrow L\pr \cr
&(1,{{{N}}}):\qquad \pl \rightarrow R\pl \qquad\qr \rightarrow R\qr .\cr}
\EQN qpal2$$
\autoparens
These quarks are also assumed to carry a color charge which will be
suppressed.  The most general $SU({{{N}}})_L\times SU({{{N}}})_R$
invariant free lagrangian one can build with $q$ and $p$ is

\offparens
$$
{\cal L}_0={\bar q}i\slashchar{\partial} q +{\bar p}i\slashchar{\partial} p
-{M_0}{\bql}\pr 
-{M_0'}{\bqr}\pl +h.c.
\EQN qpal3$$\autoparens
Parity conservation implies ${M_0}={M_0'}$ which gives

$$
{\cal L}_0={\bar q}i\slashchar{\partial} q +{\bar p}i\slashchar{\partial} p
-{M_0}({\bar q}p+{\bar p}q).
\EQN qpal3$$
This lagrangian clearly has symmetries beyond those assumed.  In order
to see the full symmetry structure it is convenient to define a new
field,

\offparens
$$\Psi=\onehtsq\left(\matrix{{q+p} \cr
                          {\gamma_5}({q-p})}\right). 
\EQN poiu$$\autoparens
In terms of this field the lagrangian, \Eq{qpal3}, becomes

$$
{\cal L}_0={\bar \Psi}i\slashchar{\partial} \Psi 
-{M_0}{\bar \Psi}\Psi .
\EQN qpal4$$
Since $\Psi$ is a $2N$-component vector, the full symmetry of the
lagrangian is $U(2{{{N}}})$. We choose to work in the original basis
because it allows a convenient identification of $U(2N)$ subgroups
with QCD symmetries\ref{beane}. So we assemble $q$ and $p$ into the
$2N$-component vector, ${\psi}=({q}\; {p})^{\sss T}$. The lagrangian,
\Eq{qpal3}, then takes the form

$$
{\cal L}_0={\bar \psi}i\slashchar{\partial} \psi 
-{M_0}{\bar \psi}{\sigma _{\sss 1}}\psi 
\EQN qpal4$$
where $\sigma _{\sss 1}$ is a Pauli matrix acting in the $q$-$p$
space. It is easy to show that in this basis the algebra of
$U(2{{{N}}})$, or $SU(2{{{N}}})\times U(1)_B$, arises from the
embedding $SU(2)\times SU({{{N}}})_V\rightarrow
SU(2{{{N}}})$\ref{beane}.  With ${{\bar\sigma} _{\sss
i}}\equiv{{\sigma} _{\sss i}}/2$, the generators in the defining
representation of $SU(2{{{N}}})$ can be written as $\{{{\bar\sigma}
_{\sss 1}}, {{\bar\sigma} _{\sss 2}}{\gamma_5}, {{\bar\sigma} _{\sss
3}}{\gamma_5}\}\otimes{\bf 1}$, ${\bf 1}\otimes{T_a}$ and
$\{{{\bar\sigma} _{\sss 1}},{{\bar\sigma} _{\sss
2}}{\gamma_5},{{\bar\sigma} _{\sss 3}}{\gamma_5}\}\otimes{T_a}$.  In
particular, ${\bf 1}\otimes{T_a}$, ${{\bar\sigma} _{\sss
3}}{\gamma_5}\otimes{T_a}$ and ${{\bar\sigma} _{\sss
3}}{\gamma_5}\otimes{\bf 1}$ are identified with $SU(N)_V$, $SU(N)_A$
and $U(1)_A$, respectively. This is consistent with the chiral
symmetry assignments in \Eq{qpal2}.

As in the $N$-flavor QCD analysis, we now add a $U(1)_A$ violating
quark interaction to take into account the effect of the axial
anomaly.  Consider the $U(1)_A$ violating, $SU({{{N}}})_L\times
SU({{{N}}})_R\times U(1)_B$ preserving 't Hooft interaction

\offparens
$$
{\cal L}''({\bar\theta} )=  
-{\kappa}\{ e^{i{\bar\theta}}\;det\; 
{\bar\psi}(1-{\sigma_{\sss 3}}{\gamma_5}){\psi} +
e^{-i{\bar\theta}}\;det\; 
{\bar\psi}(1+{\sigma_{\sss 3}}{\gamma_5}){\psi} \}, 
\EQN u1expbr$$\autoparens
where the determinant acts on $SU({{{N}}})$ matrices and $\kappa$ is a
parameter of mass dimension $4-3{{{N}}}$. We have included a P and CP
violating phase, ${\bar\theta}$, which is again assumed to contain
both pure QCD effects and quark mass matrix effects.  Note that
${\bpr}\pl + {\bql}\qr =
\oneht {\bar\psi}(1-{\sigma_{\sss 3}}{\gamma_5}){\psi}$ 
and ${\bpl}\pr + {\bqr}\ql = 
\oneht {\bar\psi}(1+{\sigma_{\sss 3}}{\gamma_5}){\psi}$. 

Although the 't Hooft interaction is constructed to taken into account
the effect of the axial anomaly, one can check that all of the
continuous subgroups of $U(2N)$ that are not identified with QCD
symmetries are broken by the 't Hooft interaction. That is, the two
$U(1)$'s, ${{\bar\sigma}_{\sss 1}}\otimes{\bf 1}$ and ${{\bar\sigma}
_{\sss 2}}{\gamma_5}\otimes{\bf 1}$, and the two $SU(N)$'s,
${{\bar\sigma}_{\sss 1}}\otimes{T_a}$ and ${{\bar\sigma} _{\sss
2}}{\gamma_5}\otimes{T_a}$ are broken by the 't Hooft interaction.
However, consider the discrete transformation $\psi\rightarrow \pm
i{\sigma _{\sss 1}}\psi$.  This transformation generates an $\discr$
($=\discrete$) subgroup of the $U(1)$ generated by ${{\bar\sigma}
_{\sss 1}}\otimes{\bf 1}$.  $\discr$, the group of permutations of two
objects, has the effect of interchanging $q$ and $p$. In the 't Hooft
interaction, $\discr$ interchanges ${\bar\psi}(1-{\sigma_{\sss
3}}{\gamma_5}){\psi}$ and ${\bar\psi}(1+{\sigma_{\sss
3}}{\gamma_5}){\psi}$, which is equivalent to the transformation
$L\leftrightarrow R$. That is

\offparens
$$
{{\cal S}_{\sss 2}}\;{\cal L}''({\bar\theta} 
)\;{{\cal S}_{\sss 2}^{-1}}=
{\cal L}''(-{\bar\theta} ).
\EQN Pinvparoof$$\autoparens
In the absence of explicit chiral symmetry breaking effects we can
perform the field redefinition, $\psi\rightarrow
e^{i{\bar\theta}{\sigma _{\sss 3}}{\gamma_5}/2N}\psi$, which removes
${\bar\theta}$ from the problem, and then P, CP {\it and} $\discr$ are
manifest discrete symmetries of the theory.

Assuming $2N$ degenerate flavors, we can include an ($\discr$
invariant) explicit chiral symmetry breaking mass term,
$-{m_q}{\bar\psi}{\psi}$, and we can again perform a field redefinition
which removes ${\bar\theta}$ from the 't Hooft interaction.  For small
${\bar\theta}$, ${\bar\psi}{\psi}\rightarrow {\bar\psi}{\psi}
+i({\bar\theta}/N){\bar\psi}{\sigma_{\sss 3}}{\gamma_5}{\psi}$, which
induces the P, CP and $\discr$ violating operator,

$$
-i{{m_q}\over N}\,{\bar\theta}\,{\bar\psi}{\sigma_{\sss 3}}{\gamma_5}{\psi}.
\EQN ebs2$$
{}From the point of view of the effective theory, this operator
contributes to the neutron electric dipole moment.  However, in
general, P, CP {\it and} $\discr$ will be broken unless
${\bar\theta}$=$0$ or $\pi$.  So in the $2N$-flavor effective theory,
${\bar\theta}$ is a natural parameter since its vanishing enlarges the
symmetry of the standard model to include $\discr$. This is to be
contrasted with the operator, \Eq{ebs2qcd}, which one has in QCD.

A natural question to ask at this point is whether one wants the
symmetry of the standard model to be enlarged to include $\discr$. In
practical terms, does $\discr$ have other consequences beyond those
discussed above?  If $\discr$ is unbroken, one would expect that the
low-energy spectrum in the $2N$-flavor effective theory in some sense
reflects $\discr$.  In the next section we will see that this is the
case.

%%%%%%%%%%%%%%%%%%%%%%%%%%%%%%%%%%%%%%%%%%%%%%%%%%%%%%%%%%%%%%%%%%%
\vskip0.1in
\noindent {\twelvepoint{\bf 4.\quad $\discr$ Doublets in the Hadron Spectrum}}
\vskip0.1in

Consider $N=2$ QCD. We have the pattern of symmetry breaking
$SU(2)_L\times SU(2)_R\rightarrow SU(2)_V$.  In the four-flavor
effective theory this will occur if the condensate
$\vev{{\bar\psi}{\psi}}$ is non-vanishing.  We will assume that
$\vev{{\bar\psi}{\sigma_{\sss 3}}{\psi}}$ vanishes since otherwise
$\discr$ is spontaneously broken.  The simplest and most convincing
way of finding consequences of $\discr$ in the low-energy theory is to
construct meson states directly out of four quarks\ref{beane}. A
general symmetry argument is given in \Ref{beane}, and an equivalent
argument using superconvergent sum rules is given in \Ref{mended}. It
is convenient to work in the diagonal basis:

\offparens
$${\phi_\pm}\equiv{\textstyle {1\over\sqrt{2}}} (q\pm p).
\EQN physeig$$ 
\autoparens
These states transform as an $\discr$ doublet with chiral
transformation properties given by \Eq{qpal2}. Since the doublet is
the only non-trivial representation of $\discr$, we expect that the
product of two doublets gives two doublets; i.e. $2\otimes 2 =2\oplus
2$. However, invariance under charge conjugation unfolds one of the
doublets since ${{\bar\phi}_+}{\phi_-}$ and ${{\bar\phi}_-}{\phi_+}$
are not states of definite charge conjugation sign.  The meson states
of definite charge conjugation sign and their associated chiral
transformation properties are:

$$\eqalign{
{\ket{\rm I}}&\sim
{{\bar\phi}^{\sss 1}_-}{\phi^{\sss 2}_-}=
\oneht ({{\bar q}_{\sss 1}}{q_{\sss 2}}+{{\bar p}_{\sss 1}}{p_{\sss 2}})-
\oneht ({{\bar q}_{\sss 1}}{p_{\sss 2}}+{{\bar p}_{\sss 1}}{q_{\sss 2}})\cr
&\quad\qquad\qquad\qquad(2,2)\qquad (1,1\oplus 3)\oplus (1\oplus 3,1)\cr
{\ket{\rm II}}&\sim
{{\bar\phi}^{\sss 1}_+}{\phi^{\sss 2}_+}=
\oneht ({{\bar q}_{\sss 1}}{q_{\sss 2}}+{{\bar p}_{\sss 1}}{p_{\sss 2}})+
\oneht ({{\bar q}_{\sss 1}}{p_{\sss 2}}+{{\bar p}_{\sss 1}}{q_{\sss 2}})\cr
&\quad\qquad\qquad\qquad(2,2)\qquad (1,1\oplus 3)\oplus (1\oplus 3,1)\cr
{\ket{\rm III}}&\sim
{{\bar\phi}^{\sss 1}_+}{\phi^{\sss 2}_-}
+{{\bar\phi}^{\sss 1}_-}{\phi^{\sss 2}_+}=
{{\bar q}_{\sss 1}}{q_{\sss 2}}-{{\bar p}_{\sss 1}}{p_{\sss 2}}\cr
&\qquad\qquad\qquad\qquad\qquad(2,2)\cr
{\ket{\rm IV}}&\sim
{{\bar\phi}^{\sss 1}_+}{\phi^{\sss 2}_-}
-{{\bar\phi}^{\sss 1}_-}{\phi^{\sss 2}_+}=
{{\bar p}_{\sss 1}}{q_{\sss 2}}-{{\bar q}_{\sss 1}}{p_{\sss 2}}\cr
&\quad\qquad\qquad\qquad(1,1\oplus 3)\oplus (1\oplus 3,1).\cr}
\EQN phenwhat$$
The numerical scripts make the permutation properties clear, and we
have used the chiral transformation properties of $q$ and $p$ given in
\Eq{qpal2}. These states have charge conjugation sign: $\pm\epsilon$ 
for ${\ket{\rm I}}$, ${\ket{\rm II}}$ and ${\ket{\rm III}}$, and
$\mp\epsilon$ for ${\ket{\rm IV}}$, and are invariant (up to a phase)
with respect to the $\discr$ transformation

$$
{q_{\sss 1}}\longleftrightarrow{p_{\sss 1}}\quad  
{q_{\sss 2}}\longleftrightarrow{p_{\sss 2}}.
\EQN condi1a$$
Note that the $\discr$ transformation

$$
{q_{\sss i}}\longleftrightarrow{p_{\sss i}}\quad\;  
{q_{\sss j}},{p_{\sss j}}\quad
{\tt fixed}\;\quad{i\neq j},
\EQN condi$$
interchanges $(2,2)$ and $(1,1\oplus 3)\oplus (1\oplus 3,1)$
representations and therefore leaves ${\ket{\rm I}}$ and ${\ket{\rm
II}}$ invariant while interchanging ${\ket{\rm III}}$ and ${\ket{\rm
IV}}$.

In the broken symmetry phase with $\vev{{\bar\psi}{\psi}}\neq 0$ it is
not generally sensible to classify states by their chiral
transformation properties, and so one might think that the chiral
decomposition of the meson states given in \Eq{phenwhat} is
useless. This is not so as there are Lorentz frames in which the
condensate decouples and the full chiral algebra is useful for
classification purposes\ref{alg}\ref{wilson}.  The infinite momentum
frame is one example of such a frame. In these Lorentz frames helicity
is conserved and so hadrons can be classified according to the full
chiral algebra for each helicity\ref{alg}. Therefore we assume that
the meson states in \Eq{phenwhat} are states of definite helicity,
parity and isospin.  Note that the insertion of additional gamma
matrices can only change the parity of the state, or interchange the
$(2,2)$ and $(1,1\oplus 3)\oplus (1\oplus 3,1)$ representations.

Consider the consequences of this multiplet structure for the ground
state of the four-flavor effective theory. The pattern of chiral
symmetry breaking determines that the lowest lying state in the
spectrum is the pion and so we are interested in the chiral
representation involving the pion.  Since the pion is a Lorentz
scalar, all states in this representation have zero-helicity. In the
case of zero-helicity there is conservation of {\it normality},
$\eta\equiv P{(-1)^J}$, where $P$ is intrinsic parity and $J$ is
spin\ref{alg}. Since $\pi$ has $\eta$=$-1$, only states of opposite
normality communicate by single-pion emission and absorption.  The
grouping we consider here is well known\ref{alg}.  The pion is joined
by a scalar $\epsilon$ ($\eta$=$+1$), and the helicity-$0$ components
of $\rho$ ($\eta$=$+1$) and ${a_{\sss 1}}$ ($\eta$=$-1$). These are
states with $GP{(-1)^J}$=$+1$ where $G$ is $G$-parity. From
\Eq{phenwhat} we identify $\ket{\rm I}$=$\ket{\pi}$,
$\ket{\rm{II}}$=$\ket{a_{\sss 1}}^{\sss (0)}$,
$\ket{\rm{III}}$=$\ket{\epsilon}$ and
$\ket{\rm{IV}}$=$\ket{\rho}^{\sss (0)}$. It is instructive to introduce
a mixing angle, $\phi$. The pion representation then takes the form:

\offparens
$$\eqalign{ 
&\ket{\pi}_a=-\cos\phi{\ket{2,2}_a}
+\sin\phi{\ket{A}_a}\cr
&\ket{a_{\sss 1}}_a^{\sss (0)}= \sin\phi{\ket{2,2}_a}
+\cos\phi{\ket{A}_a}\cr
&\ket{\epsilon}={\ket{2,2}_4}\qquad
\ket{\rho}_a^{\sss (0)}={\ket{V}_a},\cr}
\EQN sfsr1$$ 
\autoparens
where
${\ket{1,3}_a}-{\ket{3,1}_a}\equiv{\textstyle{\sqrt{2}}}{\ket{V}_a}$
and
${\ket{1,3}_a}+{\ket{3,1}_a}\equiv{\textstyle{\sqrt{2}}}{\ket{A}_a}$,
the superscripts denote helicity and the subscripts are isospin
indices.  By considering matrix elements of the vector and axialvector
currents between these states one finds that
$f_\pi$=$f_{\rho}\sin\phi$ and $f_{a_{\sss
1}}$=$f_{\rho}\cos\phi$\ref{beane}. In helicity conserving frames the
mass-squared matrix is a relevant quantity. All mass-squared
splittings transform like $\vev{{\bar\psi}\psi}$; i.e., like the
fourth component of a chiral four-vector\ref{alg}\ref{beane}. Setting
$M_\pi^2$=$0$ one then obtains ${M_\rho^2}=\cos^2\phi{M_{a_{\sss
1}}^2}$. By considering matrix elements of the pion transition
operator, $X_a^\lambda$, one also finds
${g_{\rho\pi\pi}^2}{f_\pi^2}={M_\rho^2}\sin^2\phi$\ref{alg}.  One can
then obtain combinations of masses and decay constants that are
independent of the mixing angle. In particular it is clear that

\offparens
$$\EQNalign{ 
&\;\;{f_{a_{\sss 1}}^2}+{f_\pi^2}={f_\rho^2} \EQN sfsr4;a\cr
&{M_\rho^2}{f_\rho^2}={M_{a_{\sss 1}}^2}{f_{a_{\sss 1}}^2}\EQN sfsr4;b\cr}
$$\autoparens which is precisely the content of the first and second
spectral function sum rules\ref{sfsr}, respectively, evaluated in
resonance saturation approximation. 

The $\discr$ invariance implies that the irreducible chiral
representations must enter with equal weight (see \Eq{condi}) and so $\pi$
and $a_{\sss 1}^{\sss (0)}$ form an $\discr$ doublet,
$\cos\phi$=$\sin\phi$=$1/\sqrt{2}$, and we obtain the familiar
KSRF relations\ref{ksrf}

\offparens
$$
{{M_\rho^2}\over{g_{\rho\pi\pi}^2}{f_\pi^2}}=2\qquad (1.89\pm 0.07)\qquad\quad
{{{f_{\rho}}{M_\rho}}\over{{g_{\rho\pi\pi}}{f_\pi^2}}}
=2\qquad (2.28\pm 0.05),
\EQN sfsr6$$ 
which are in very good agreement with the experimental numbers (in
parentheses) extracted from the decays ${\rho^{\sss
0}}\rightarrow{\pi^+}{\pi^-}$ and ${\rho^{\sss
0}}\rightarrow{e^+}{e^-}$\ref{pdg}.
\Eq{sfsr6} in turn implies ${f_{a_{\sss 1}}^2}={f_\pi^2}$,
$2{M_\rho^2}={M_{a_{\sss 1}}^2}$ and ${M_\epsilon^2}={M_\rho^2}$. We
emphasize that these relations, which have been derived previously
by imposing asymptotic constraints on pion-hadron scattering
amplitudes\ref{alg}, are exact consequences of $\discr$ in the
four-flavor effective theory.

On the basis of quark model intuition one might have erroneously
supposed that the four-flavor effective theory would give a
pathological spectrum. The effective theory gets the ground state
correctly because it respects ---by assumption--- the pattern of
chiral symmetry breaking.  From the quark point of view chiral
symmetry breaking generates a mass gap which splits the four
degenerate flavors so that two remain massless ---giving the pion pole
and a conserved axial current--- and two become massive\ref{beane}.

%%%%%%%%%%%%%%%%%%%%%%%%%%%%%%%%%%%%%%%%%%%%%%%%%%%%%%%%%%%%%%%%%%%
\vskip0.1in
\noindent {\twelvepoint{\bf 5.\quad Conclusion}}
\vskip0.1in

We have considered an effective theory of QCD with twice the number of
QCD flavors. When the axial anomaly is taken into account this
effective theory has the same continuous global symmetries as QCD.
However, there is a residual discrete symmetry that has several
interesting consequences.  In particular, if this $\discr$ symmetry is
unbroken, there is no P or CP violation in the effective theory.  One
might conclude that since $\bar\theta$ is an unconstrained parameter
in QCD, the effective theory cannot be describing the same physics as
QCD. This viewpoint can be questioned because $\discr$ has additional
consequences in the effective theory.  In particular, one can show
that there is a sense in which hadrons in the effective theory are in
$\discr$ multiplets, leading to predictions in agreement with
experiment and which have been found previously by assuming
superconvergent sum rules in pion-hadron scattering\ref{mended}.  Of
course, if $\discr$ is a global symmetry one does not expect it to be
exact and so it can clearly be broken in a manner which preserves the
predictions of section 4 and yet grossly violates the experimental
bound on $\bar\theta$.

One interesting possibility is that $\discr$ is a discrete gauge
symmetry\ref{krauss}.  If $\discr$ is a gauge symmetry, then the
$2N$-flavor effective theory has the same global symmetries as
$N$-flavor QCD.  In this case, $\discr$ seems to solve the strong CP
problem without conflicting with expectations that global symmetries
are sacred and should therefore be shared by different descriptions of
the same physics. This interpretation is consistent with the fact that
$\discr$ only seems to have consequences related to asymptotic
behavior of scattering amplitudes.  Although gauge symmetries are
redundancies, they do have algebraic consequences when married with
asymptotic constraints. In this sense gauge symmetries behave like
spontaneously broken chiral symmetries\ref{alg}. A nice example is
that of the Drell-Hearn-Gerasimov sum rule which can be expressed
algebraically as a statement of the (trivial) $U(1)$ algebra of
electromagnetism (see the fourth entry in
\Ref{alg} and also \Ref{mended}).

We should also note that the idea of parity doublets arising
non-perturbatively in the infrared as rescuers of discrete space-time
symmetries has recently been considered in \Ref{bala}.

\vskip0.15in
\noindent
This work was supported by the U.S. Department of Energy (Grant
DE-FG05-90ER40592 at Duke and grant DE-FG02-93ER-40762 at Maryland). I
thank T.D.~Cohen, M.A.~Luty, M.~Malheiro and B.~M\"uller for valuable
conversations and criticism.

%\vfill\eject % new page 
\nosechead{References}% % no section number
%\addTOC{1}{References}{\folio}% % add to contents
%\global\def\HeadText{{\tenit References}}% % running head text
\ListReferences \vfill\supereject \end